\documentstyle[12pt]{article}
\begin{document}
\baselineskip=20pt
\pagestyle{myheadings}
\noindent
\centerline{{{\large\bf Thomas Rotation and Polarised Light: }}}
\centerline{{{\large\bf A non-Abelian Geometric Phase in Optics}}}
\bigskip
\bigskip
\centerline{Joseph Samuel and Supurna Sinha }
\centerline{Raman Research Institute}
\centerline{Bangalore, India 560 080}
\bigskip
\bigskip
\begin{abstract}
We describe a {\it non-Abelian} Berry phase in
polarisation optics, suggested by an analogy due to Nityananda
between boosts in special relativity and the effect of elliptic
dichroism on polarised light. The analogy permits a simple 
optical realization of the non-Abelian gauge field 
describing Thomas rotation.
We also show how Thomas rotation can be understood geometrically
on the Poincar\'{e} sphere 
in terms of the Pancharatnam phase.
\end{abstract}
\vspace*{6cm}
To appear in Pramana
\bigskip
\newpage
\section{Introduction}
Shortly after Berry's discovery of the geometric phase
\cite{Berry}, the idea was generalized in various directions 
\cite{Wil,AA,Gene,Book}. 
Here we follow on the work of Wilczek and Zee \cite{Wil},
who noted that in quantum systems with N-fold degeneracy, the
``phase" is no longer Abelian, but an element of $U(N)$,
the unitary group in $N$ dimensions. These non-Abelian phases
have been studied in 
Nuclear Quadrupole resonance (NQR) \cite{Book}. 

There have been several demonstrations of the geometric phase
\cite{Bh,Si,Sim,Ha,Berk,Hari} in 
polarisation optics. However, {\it all} of these experiments
deal with the {\it Abelian} phase that Berry's work drew wide
attention to. The purpose of this paper is to point out that the
{\it non-Abelian} phase of Wilczek and Zee \cite{Wil} can also be
realised in polarisation optics. Since optical systems are much easier
to set up than say, NQR experiments, we feel that our
observation will aid 
experimental studies of the non-Abelian phase.

There has been some interest in the gauge theoretic 
aspects of Thomas rotation \cite{Shankar,Mer}, which 
can be abstractly \cite{Mer}
viewed as an $SO(3)$ gauge field living on the space of inertial 
observers.
More concretely, Mathur \cite{Shankar} pointed out 
that the Thomas precession gauge field (TPGF) appears as a non-Abelian
Berry phase due to Kramer's degeneracy of the Dirac electron. 
Our purpose in this article is to present another concrete realization
of the TPGF in 
polarisation optics. We believe our work may be of general interest,
as it permits a simple optical demonstration of Thomas
Rotation (TR). While our study is both theoretical and experimental, this
paper is devoted
to theoretical aspects. Experimental details will be 
reported elsewhere.

In section 2 we review the gauge theoretic aspects of Thomas Rotation
as a prelude to realising it in optics in section 3. In section
4 we describe how the phenomenon of TR can be understood in {\it
purely optical terms} using the Pancharatnam Phase. Section 5 is
a concluding discussion.
\section{Thomas Rotation}
As is well known in the special theory of relativity,
{\it pure} 
boosts (changes of frame without rotation of the spatial 
axes)
along different spatial directions do not commute with each other.
In fact, the commutator of two infinitesimal boosts is a rotation and
this leads to Thomas precession
\cite{Thomas,Jackson}.
We deal here with the {\it finite}
version of this effect, the Thomas Rotation \cite{Urb,Mer}. 
Let $u_1^\mu, (\mu=0,1,2,3)$ be the four-velocity
of an observer and $\Lambda_{21},\Lambda_{32}$ and $\Lambda_{13}$ a
sequence of pure boosts ($\Lambda$ is given explicitly in \cite{Mer}) 
applied to $u_1$. The sequence of 
pure boosts takes $u_1$ to $u_2$ ($=\Lambda_{21}u_1$),
$u_3$ ($=\Lambda_{32}u_2$) and finally back to $u_1$ ($=\Lambda_{13}u_3$). 
Since the final four-velocity $\Lambda_{13}\Lambda_{32}\Lambda_{21}
u_1$ is equal to the initial four-velocity $u_1$,
we say that the sequence of boosts 
{\it closes}   
(returns the observer to her original frame). In general, the product 
$\Lambda = \Lambda_{13}\Lambda_{32}\Lambda_{21}$ of
boosts is not the identity, but $\Lambda = R$,
where $R$ is a rotation matrix representing the Thomas rotation.
The rotation matrix $R$ can be expressed as a path ordered
product or Wilson loop:
\begin{equation}
R=Pexp[\oint A(u)du],
\label{Wilson}
\end{equation}
where the integral is over the geodesic triangle $u_1-u_2-u_3-u_1$
on the unit hyperboloid ${\mbox{${\cal H}^{+}$}}$ 
of four velocities ${\mbox{${\cal H}^{+}$}}=\{u^\mu|u.u=1,\, u^0> 0\}$.
The gauge field $A$ describing Thomas rotation is a $3\times3$
antisymmetric matrix, whose
components (in a convenient \cite{Mer} global gauge choice over 
${\mbox{${\cal H}^{+}$}}$) are
$A^{ i}{}_{ j}=(1+u^0)^{-1}(du^i u_j-du_j u^i)$, where $i,j=1,2,3$.
A closed sequence of three boosts must lie in a plane \cite{Mer}.
This is no longer true for a closed sequence of $M$ (four or more)
boosts, which involves $M$ inertial observers with $M$ four-velocities
$u_1,u_2,...u_M$. However, one can consider these $M$ observers in triplets
by triangulating the broken geodesic curve $C=u_1-u_2-...u_M-u_1$
and applying the previous argument. The Thomas rotation is then
expressed as a Wilson loop over $C$. In general $C$ is a 
(closed, piecewise geodesic) {\it space} curve in ${\mbox{${\cal H}^{+}$}}$
and path ordering in (\ref{Wilson}) becomes necessary, revealing the 
essentially {\it non-Abelian} nature of the gauge field. One can also consider
the limit $M\rightarrow\infty$ of an infinite number of infinitesimal
boosts. The integral in (\ref{Wilson}) is then over a continuous
closed curve in ${\mbox{${\cal H}^{+}$}}$, which is in general nonplanar.
Each boost distorts (alters the distances between points on)
the celestial sphere of an observer
due to the well known phenomenon of aberration of light 
\cite{Jackson,Rindler} from
the sky. When the sequence of boosts closes, the distortions cancel
and the net result is an undistorted but rotated celestial
sphere, the rotation matrix being given by (\ref{Wilson}). 
\section{Analogue of Thomas Rotation in Optics}
In order to realise the TPGF $A^i_j$ in optics, 
we draw on an observation by Nityananda \cite{RN} 
regarding the
effect of an anisotropic absorber on the polarisation 
of light. 
Let ${\mbox{${\hat{n}}$}}$ be a unit vector representing a general point on
the Poincar\'{e} sphere. Suppose that light is passed through an
absorber  $A_{{\mbox{${\hat{n}}$}}}$
that preferentially absorbs the polarisation state ${\mbox{${\hat{s}}$}}$ orthogonal to
${\mbox{${\hat{n}}$}}$. 
If the incident light is in the state ${\mbox{${\hat{n}}$}}$ or ${\mbox{${\hat{s}}$}}$, its
polarisation state will not be affected by the material (although its
intensity may diminish).  
Any other polarisation state ${\mbox{${\hat{p}}$}}$ on the Poincar\'{e} sphere 
can be resolved in terms of
${\mbox{${\hat{n}}$}}$ and ${\mbox{${\hat{s}}$}}$ and, since the  ${\mbox{${\hat{s}}$}}$ component of ${\mbox{${\hat{p}}$}}$ 
is preferentially absorbed,
will
move towards ${\mbox{${\hat{n}}$}}$ along the great circle \cite{RSGNR} joining 
${\mbox{${\hat{p}}$}}$ to ${\mbox{${\hat{n}}$}}$ . The key observation due to 
Nityananda \cite{RN} 
is that the effect
of an absorber $A_{{\mbox{${\hat{n}}$}}}$  on the Poincar\'{e} sphere is identical to
the effect of a pure Lorentz boost in the direction ${\mbox{${\hat{n}}$}}$ 
on the celestial sphere of an
observer in special relativity. 

The optical element $A_{{\mbox{${\hat{n}}$}}}$
discussed above is purely absorbing (dichroic) in the sense 
that it introduces only a relative attenuation between two orthogonal
states and not a relative phase. Dichroic 
elements $A_{{\mbox{${\hat{n}}$}}}$ distort the 
Poincar\'{e} sphere. There are also birefringent elements
$R_{{\mbox{${\hat{n}}$}}}$ (retarders, such as wave plates), 
which introduce a
{\it pure} phase difference between two orthogonal states,
and rigidly rotate the Poincar\'{e} sphere about an axis
${\mbox{${\hat{n}}$}}$. The effect of absorbers 
and retarders on the Poincar\'{e}
sphere is qualitatively different.
Yet, by exploiting the analogy
with special relativity, we see that a sequence of absorptions 
along different directions can give rise to a net rigid rotation
of the Poincar\'{e} sphere. 
A sequence of elliptic dichroids can result
in a net elliptic birefringence. {\it This is the optical
analogue of Thomas rotation}.

Before developing this analogy in quantitative terms, 
we briefly clarify our terminology, which some readers may find unfamiliar.
Traditionally, optical elements which cause a rotation of the
Poincar\'{e} sphere about the x or y axis are called birefringent, whereas
those which generate a rotation about the z axis are called
optically active. There is no standard terminology
for an element which causes a rotation of the Poincar\'{e} sphere about
an arbitrary axis. 
Since these are merely rotations about
different axes, there is no fundamental difference between them.
We adopt a generalised terminology and 
refer to all of these as elliptically birefringent elements. 
Special cases of elliptic birefringence are linear (ordinary birefringence),
and circular (optical activity). We also use the word dichroism
in the same general sense to refer to elliptic dichroism and not
just linear or circular dichroism.

Consider an absorbing optical 
device $A_{{\mbox{${\hat{n}}$}}}$ whose effect on a state $|n>$
is to reduce its amplitude to $e^{-\alpha_{1}}|n>$
and whose effect on the state $|s>$ orthogonal to $|n>$ 
is to reduce its amplitude to $e^{-\alpha_{2}}|s>$
(we assume $\alpha_{2} > \alpha_{1}$, so state $|s>$ is preferentially
absorbed). 
It is convenient to introduce the relative and overall absorption
coefficients $\alpha=\alpha_2-\alpha_1$ and $\alpha_0=(\alpha_1+\alpha_2)/2$
respectively.  Since we are not interested in the overall intensity and
phase of the light beam (these are not represented on the Poincar\'{e}
sphere), we do not need to normalise our state vectors. 
A general (unnormalised) state $|p>$ can be expanded in the 
orthonormal basis $\{|n>, |s>\}$ : $|p> = |n> + z|s>$, 
where $z = \tan(\theta /2)e^{i\phi}$ with $\theta$ the angle between 
${\mbox{${\hat{p}}$}}$ and ${\mbox{${\hat{n}}$}}$ on the Poincar\'{e} sphere (the ``colatitude'')
and $\phi$ the ``longitude".
The effect
of $A_{{\mbox{${\hat{n}}$}}}$ on $|p>$ is to transform it to 
$|p'> = |n> + e^{-\alpha} z|s>$ (where we have discarded
an overall factor which does not affect the polarisation state).
On the Poincar\'{e} sphere ${\mbox{${\hat{p}}'$}}$ has the same ``longitude''
as ${\mbox{${\hat{p}}$}}$ and makes an angle 
$\theta'$ with ${\mbox{${\hat{n}}$}}$ where $\tan(\theta' /2) = e^{-\alpha} \tan(\theta /2)$.
A little algebra shows that $\cos \theta'=(\cos \theta+\tanh
\alpha)/(1+\cos \theta \tanh \alpha)$. Letting $\beta=\tanh \alpha$,
we notice that this is identical to the aberration formula \cite{Jackson}. 
The relative absorption $\alpha$ plays the role of rapidity in
the analogue relativistic system and $\beta$ is the velocity in
units of the speed of light \cite{fo1}. An absorber
$A_{{\mbox{${\hat{n}}$}}}(\alpha)$ with
relative absorption $\alpha$ corresponds to a boost
$\Lambda_{{\mbox{${\hat{n}}$}}}(\alpha)$ in the ${\mbox{${\hat{n}}$}}$ direction with rapidity $\alpha$.
By applying successive boosts $\Lambda_{{\mbox{${\hat{n}}$}}}(\alpha)$ to the initial
four-velocity $u_1^\mu$ (say $(1,0,0,0)$), one can trace the path of the
equivalent relativistic system \cite{fo2}
on ${\mbox{${\cal H}^{+}$}}$. 
\section{Optical View of Thomas Rotation}
We now show how Thomas rotation can be understood in purely
optical terms, using the Pancharatnam phase \cite{Panch,RSRN,Berj}.
Let a light beam pass through a sequence of three absorbers 
$A_{{\mbox{${\hat{n}}$}}_1}, A_{{\mbox{${\hat{n}}$}}_2}$ and $A_{{\mbox{${\hat{n}}$}}_3}$ which closes (the combination
$A=A_{{\mbox{${\hat{n}}$}}_3}A_{{\mbox{${\hat{n}}$}}_2}A_{{\mbox{${\hat{n}}$}}_1}$ preserves distances between points on
the Poincar\'{e} sphere).
Consider the set
of points ${\mbox{$\cal C$}}$ on the great circle through ${\mbox{${\hat{n}}$}}_1$ and ${\mbox{${\hat{n}}$}}_2$. Since these
points are ``dragged'' {\it along} ${\mbox{$\cal C$}}$ by both $A_{{\mbox{${\hat{n}}$}}_1}$ and $A_{{\mbox{${\hat{n}}$}}_2}$, the 
set ${\mbox{$\cal C$}}$ is left invariant by both these absorbers.\\
{\it Lemma 1:} If the sequence $A_{{\mbox{${\hat{n}}$}}_1},A_{{\mbox{${\hat{n}}$}}_2},A_{{\mbox{${\hat{n}}$}}_3}$ closes,
${\mbox{${\hat{n}}$}}_3$ must lie on the great circle ${\mbox{$\cal C$}}$.\\
{\it Proof:} If ${\mbox{${\hat{n}}$}}_3$ is not on ${\mbox{$\cal C$}}$, it must lie on one of the
hemispheres into which ${\mbox{$\cal C$}}$ divides the Poincar\'{e} sphere.
$A_{{\mbox{${\hat{n}}$}}_1}$ and $A_{{\mbox{${\hat{n}}$}}_2}$ leave ${\mbox{$\cal C$}}$ invariant, but the effect of $A_{{\mbox{${\hat{n}}$}}_3}$
is to drag all points of ${\mbox{$\cal C$}}$ into the hemisphere containing ${\mbox{${\hat{n}}$}}_3$.
Antipodal points of ${\mbox{$\cal C$}}$ (which are separated by a distance $\pi$) will
no longer be antipodal since they are in the same hemisphere. This contradicts
the assumption that the sequence of three absorbers closes. It follows that
the points ${\mbox{${\hat{n}}$}}_1,{\mbox{${\hat{n}}$}}_2,{\mbox{${\hat{n}}$}}_3$ lie on a great circle.\\
${\mbox{$\cal C$}}$ is left invariant by all the three absorptions and therefore by their
composition $A=A_{{\mbox{${\hat{n}}$}}_3}A_{{\mbox{${\hat{n}}$}}_2}A_{{\mbox{${\hat{n}}$}}_1}$. Let ${\mbox{${\hat{n}}$}}$ and ${\mbox{${\hat{s}}$}}$ be the two
points on the Poincar\'{e} sphere which satisfy
${\mbox{${\hat{n}}$}}.{\mbox{${\hat{n}}$}}_i={\mbox{${\hat{s}}$}}.{\mbox{${\hat{n}}$}}_i=0$ for $i=1,2,3$.\\
{\it Lemma 2:} ${\mbox{${\hat{n}}$}}$ and ${\mbox{${\hat{s}}$}}$ are returned to their original positions by
the sequence $A=A_{{\mbox{${\hat{n}}$}}_3}A_{{\mbox{${\hat{n}}$}}_2}A_{{\mbox{${\hat{n}}$}}_1}$.\\
{\it Proof:} Since ${\mbox{$\cal C$}}$ is invariant under 
the action of $A_{{\mbox{${\hat{n}}$}}_3},A_{{\mbox{${\hat{n}}$}}_2}$ and $A_{{\mbox{${\hat{n}}$}}_1}$,
points on the Poincar\'{e} sphere do not cross ${\mbox{$\cal C$}}$.
Since ${\mbox{${\hat{n}}$}}$ is 
the only point on its hemisphere which is $\pi/2$ away
from ${\mbox{$\cal C$}}$, ${\mbox{${\hat{n}}$}}$ must be mapped to itself by $A$. Similarly,
${\mbox{${\hat{s}}$}}$, which is antipodal to ${\mbox{${\hat{n}}$}}$, is also mapped to itself.\\
{\it Lemma 3:} The action of $A$ on the Poincar\'{e} sphere
is a rotation about the ${\mbox{${\hat{n}}$}}-{\mbox{${\hat{s}}$}}$ axis.\\
{\it Proof:} Under the action of $A_{{\mbox{${\hat{n}}$}}_3}A_{{\mbox{${\hat{n}}$}}_2}A_{{\mbox{${\hat{n}}$}}_1}$,
${\mbox{${\hat{n}}$}}$ and ${\mbox{${\hat{s}}$}}$ are dragged along great circles towards ${\mbox{${\hat{n}}$}}_1,{\mbox{${\hat{n}}$}}_2$
and ${\mbox{${\hat{n}}$}}_3$ as shown in Fig.1. (The closure of this sequence
of geodesic arcs is guaranteed by {\it Lemma 2}).
From Pancharatnam's theorem \cite{Panch,RSRN}, we see that the state ${\mbox{${\hat{n}}$}}$
picks up a geometric phase $\Omega/2$, where $\Omega$ is the solid
angle subtended by the geodesic triangle traced by the state ${\mbox{${\hat{n}}$}}$ as it
traverses the geodesic arcs. From Fig.1, note that states 
${\mbox{${\hat{n}}$}}$ and 
${\mbox{${\hat{s}}$}}$ pick up
equal and opposite geometric phases $\pm\Omega/2$. It follows
from the invariance of ${\mbox{$\cal C$}}$ that $A$ introduces no
relative attenuation between states ${\mbox{${\hat{n}}$}}$ and 
${\mbox{${\hat{s}}$}}$. This is the signature of elliptic
birefringence in the ${\mbox{${\hat{n}}$}}$- 
${\mbox{${\hat{s}}$}}$ direction.
The effect of $A$ on the Poincar\'{e} sphere is a rotation
through an angle $\Omega$ about
the ${\mbox{${\hat{n}}$}}-{\mbox{${\hat{s}}$}}$ axis. 
It is elementary to verify that 
$\Omega$ is equal to the Thomas rotation angle \cite{Urb}.

It now follows that the rotation of the Poincar\'{e} sphere due
to a closed sequence of $M$ absorbers is given by the Wilson
loop (\ref{Wilson}) over the curve $C$.
A sequence of $M$ absorbers can be represented
as a piecewise geodesic closed space curve $C$ in ${\mbox{${\cal H}^{+}$}}$. By triangulation,
one can reduce the traversal of $C$ into a sequence of traversals of triangles.
Each triangle causes a rotation about some axis as discussed
above in {\it Lemma 3}. 
Since $C$ is not in general planar, the rotations caused
by different triangles are about {\it different} axes. As a result,
one has to {\it compose} the sequence of rotations (and not just
add the rotation angles, as one does in the Abelian case). The final
answer is a path ordered product of individual rotations (\ref{Wilson}).
\section{Conclusion}
A particular case of the optical effect discussed here has been 
noticed earlier: linear
dichroism can lead \cite{Rang,Kitano} to 
optical activity (circular birefringence). 
Kitano et al \cite{Kitano} study Lorentz group Berry phases, but
since they use only {\it linear absorbers}, they only explore an
{\it Abelian} phase. As Zee \cite{Zee} pointed out in the context of the
NQR experiment of Tycko \cite{Tycko}, this amounts to
exploring a one parameter {\it Abelian subgroup} of the full
{\it non-Abelian
gauge group}. 
There is also a difference between our approach and Ref.\cite{Kitano},
which views the linear absorbers as ``squeezing" the polarisation
ellipse  similar to Berry's
discussion of the 
time dependent oscillator \cite{BerryJphys}.
The relevant group is then $SO(2,1)$ and the Abelian phase,
a $U(1)$ phase. Our approach is based on {\it distortions of the 
Poincar\'{e} sphere}. The relevant group here is $SO(3,1)$, the 
Lorentz group in $3+1$ dimensions, which results in a 
{\it Non-Abelian} gauge group $SO(3)$.

We have dealt with purely absorbing elements giving rise to a net
birefringence. In general optical elements will have both absorption
and birefringence. In the language used in the Berry phase literature,
the birefringence is the dynamic ``phase" and has to be ``subtracted out"
before one sees the geometric ``phase". In our exposition we chose
optical 
elements which are purely absorbing so that the dynamic ``phase" is 
zero. This is similar in spirit to recent demonstration of the 
Pancharatnam phase from pure projections \cite{Berk,Hari} where
the dynamic phase is manifestly zero. Indeed, in the limit of
infinite relative absorption $(\alpha \rightarrow \infty)$ absorbers
become projectors. It should be 
borne in mind that this limit is highly singular,  
corresponding in the relativistic language, to boosting a frame
to the speed of light.

There is a mathematical similarity between 
the effect described here 
and the Pancharatnam effect. Both 
demonstrate the mathematical phenomenon of anholonomy of which
there are numerous instances in physics \cite{Book}.
In both cases, the mathematical structure explored is a connection
on a fiber bundle $({\mbox{$\cal P$}},{\mbox{$\cal B$}},\Pi)$, where ${\mbox{$\cal P$}}$ is the total space,
{\mbox{$\cal B$}}, the base space and $\Pi$ a projection from ${\mbox{$\cal P$}}$ to ${\mbox{$\cal B$}}$. $\Pi^{-1}(b)$
is called the fiber over $b$. A connection on a fiber bundle
gives a rule for comparing points on neighboring fibers.
In general, the connection is not integrable. When one parallel transports
a point $p\in \Pi^{-1}(b)$ along a 
closed curve in ${\mbox{$\cal B$}}$, one returns to the same fiber,
but not in general to the same point on that fiber. 
This is a reflection of the curvature of the connection or the nonintegrability
of the rule for comparison of points. In the context of the 
Pancharatnam phase, the base space ${\mbox{$\cal B$}}$ is the Poincar\'{e} 
sphere $S^2$, the fiber is the
phase $U(1)$ and the total space ${\mbox{$\cal P$}}$ is  $S^3$. In the case of present
interest, the base space is the unit hyperboloid ${\mbox{${\cal H}^{+}$}}$ of four-velocities
and the fiber is the rotation group $SO(3)$ and the total space
is the Lorentz Group. 

The optical realization of TPGF is a simple analogue experiment
which can be easily performed. In the Dirac electron realization
of TPGF \cite{Shankar}, 
relativistic effects like TR are small under normal laboratory
conditions. In the analogue optical system, 
attaining ``relativistic" speeds
is quite easy. One just uses an absorber with
a high relative absorption. One is limited only by one's ability
to detect the residual intensity of preferentially absorbed polarisation.
In preliminary studies, we have measured
Thomas rotation angles of upto $50^{0}$ in the optics
laboratory. A possible application of the ideas described in this
paper is the design of achromatic (wavelength independent) 
birefringent elements. Because of
the geometric character of the effect, birefringent devices
(optical rotators or wave plates) constructed from dichroic
elements as described above will introduce 
the same phase difference independent of
wavelength, provided of course, the dichroism is not wavelength
dependent in the frequency bandwidth of interest.

Historically, polarised light has played a significant role 
in elucidating the field of Geometric phases. We believe 
our paper goes further in this direction by showing how 
{\it Non-Abelian} phases can also be seen in optics.

{\it Acknowledgements:} It is a pleasure to thank Rajaram
Nityananda, V.A. Raghunathan and T.N. Ruckmongathan for discussions and
N. Narayanaswamy, V. Venu, Manohar Modgekar and C.Md. 
Ateequllah for technical support in the laboratory. 
We thank Rajendra Bhandari for drawing
our attention to Ref. \cite{Kitano}.

\newpage
\section*{Figure Caption}

\noindent 
Figure 1 : The Poincar\'{e} sphere representation of the trajectories
of the two orthogonal elliptic polarisations ${\mbox{${\hat{n}}$}}$ and ${\mbox{${\hat{s}}$}}$.
These trajectories are mirror images of each other in the plane
containing ${\mbox{${\hat{n}}$}}_1,{\mbox{${\hat{n}}$}}_2$ and ${\mbox{${\hat{n}}$}}_3$. 
Note that by Pancharatnam's theorem,
these two polarisations
pick up equal and opposite geometric
phases $\pm\Omega/2$ . This 
corresponds to a rotation of the Poincar\'{e} sphere
by an angle $\Omega$ about the axis passing through the
orthogonal elliptic states ${\mbox{${\hat{n}}$}}$ and ${\mbox{${\hat{s}}$}}$.

\vskip .21cm
\noindent 

\newpage
\begin{thebibliography}{99}
\small
\setlength{\itemsep}{-0.9\parsep}
\bibitem{Berry}
M.V.Berry, {\it Proc. Roy. Soc. A}, {\bf 392}, 45 (1984).
\bibitem{Wil}
F.Wilczek and A.Zee, Phys. Rev. Lett. {\bf 52}, 2111 (1984).
\bibitem{AA}
Y.Aharonov and J.Anandan {\it Phys. Rev. Lett.} 
{\bf 58}, 1593 (1987).
\bibitem{Gene}
J.Samuel and R.Bhandari, Phys. Rev. Lett., {\bf 60}, 2339 (1988).
\bibitem{Book}
A.Shapere and F.Wilczek ``Geometric Phases in Physics'',
World Scientific, Singapore (1989).
\bibitem{Bh}
R. Bhandari and J. Samuel, 1988,
Phys. Rev. Lett., {\bf 60}, 211.
\bibitem{Si}
R. Simon, H.J. Kimble and E.C.G. Sudarshan, 1988,
Phys. Rev. Lett., {\bf 61}, 19. 
\bibitem{Sim}
T.H. Chyba, L.J. Wang, 
L. Mandel and R. Simon, 1988, Opt. Lett., {\bf 13} 562.
\bibitem{Ha}
P. Hariharan and M. Roy, 1992, J. Mod. Opt., {\bf 39}, 1811.
\bibitem{Berk}
M.V.Berry and S.Klein, 1996, J. Mod. Opt., {\bf 43}, 165.
\bibitem{Hari}
P.Hariharan, Hema Ramachandran, K.A.Suresh and J.Samuel,
``The Pancharatnam Phase as a strictly geometric phase:
a demonstration using pure projections'' RRI-96-18,
To appear in the Journal of Modern Optics.
\bibitem{Shankar}
H.Mathur, Phys. Rev. Lett. {\bf 67}, 3325 (1991).
R.Shankar and H.Mathur, Phys. Rev. Lett. {\bf 73}, 1565 (1994).
\bibitem{Mer}
J.Samuel, Phys. Rev. Lett. {\bf 76}, 717 (1996).
\bibitem{Thomas}
L.H.Thomas, Phil. Mag. {\bf 3}, 1 (1927).
\bibitem{Jackson}
J.D. Jackson ``Classical Electrodynamics'' 
Wiley Eastern Limited, New Delhi (1978).
\bibitem{Urb}
H.Urbantke, Am. J. Phys. {\bf 58}, 747 (1990).
\bibitem{Rindler}
R.Penrose and W.Rindler, ``Spinors and Spacetime, Vol I'',
Cambridge University Press, Cambridge (1984).
\bibitem{RN}
See the remark attributed to R.Nityananda 
on page 245 in the article by S.Ramaseshan in Essays on Particles
and Fields, Edited by R.R.Daniel and B.V.Sreekantan (Indian Academy
of Sciences, Bangalore (1989)).
\bibitem{RSGNR}
G.N.Ramachandran, and S.Ramaseshan,`` Crystal Optics '' in 
Handbuch der Physik, Vol. 25, Part I, Springer, Berlin (1961).
\bibitem{fo1}
In passing, we note that the 
aberration formula is considerably simplified by using 
an angular variable $\sigma$ defined by
$\cos \theta= \tanh \sigma$. The
aberration formula then reads $\sigma'=\sigma+\alpha$.
\bibitem{fo2}
If one forms the (impure)
density matrix $\rho=u^0+{\vec u}.{\vec \sigma}$, the components
$(u^0,{\vec u})$ have an optical interpretation in terms of the
Stokes \cite{RSGNR} parameters $(I,Q,U,V)$ of a partially
polarised light beam described by
that density matrix.
\bibitem{Panch}
S.Pancharatnam, {\it Proc. Indian Acad. Sci. A} 
{\bf 44}, 247 (1956), reprinted in 
``Collected works of S.Pancharatnam'' Oxford University
Press, London (1975).
\bibitem{RSRN}
S.Ramaseshan and R.Nityananda, {\it Current Sci.} (India),
{\bf 55}, 1225 (1986).
\bibitem{Berj}
MV. Berry, J. Mod. Opt. {\bf 34}, 1401 (1987).
\bibitem{Rang}
G.S.Ranganath, K.A.Suresh, S.R.Rajagopalan  and U.D.Kini,
Pramana Special Supplement {\bf 1}, 353 (1975).
\bibitem{Kitano}
M. Kitano and T. Yabuzaki, Phys. Lett  {\bf A 142}, 321  (1989).
\bibitem{Zee}
A.Zee, Phys. Rev. {\bf A 38}, 1 (1988).
\bibitem{Tycko}
R.Tycko Phys. Rev. Lett. {\bf 58}, 2281 (1987).
\bibitem{BerryJphys}
M.V. Berry, J. Phys. {\bf A 18} 15 (1985).
\end{thebibliography}
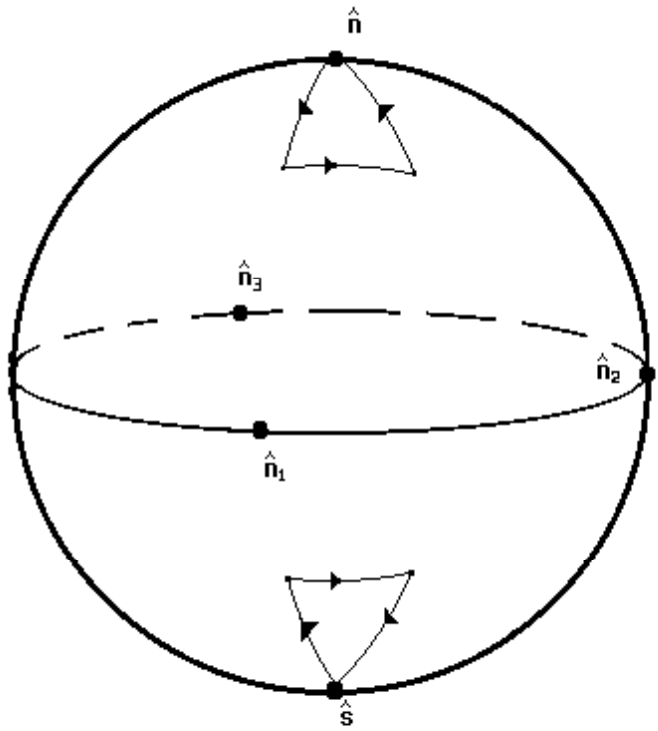
\end{document}